\documentclass[12pt]{article}
\usepackage[T2A]{fontenc}
\usepackage[cp1251]{inputenc}
\usepackage{graphicx}
\newcommand{\be}{\begin{equation}}
\newcommand{\ee}{\end{equation}}
\begin{document}
\large
\baselineskip=20pt
\centerline{Printed in: \bf Astronomy Report, 2003, v.47, p.186}
\bigskip

\centerline{\bf
Supersoft
X-ray
Sources.
Parameters
of
Stellar
Atmospheres}
\normalsize

\begin{center}
A.
A.
Ibragimov$^1$,
V.F.Suleimanov$^1$,
A.
Vikhlinin$^2$,
and
N.
A.
Sakhibullin$^1$
\\
{\it 1 - Kazan
State
University,
Kazan,
Tatarstan,
Russia}\\
2 - {\it Space
Research
Institute,
Moscow,
Russia}
\\
Received
March
5,
2002;
in
final
form,
October
10,
2002
\end{center}

\bigskip

\centerline {\bf Abstract}

ROSAT
spectra
of
11
supersoft
X-ray
sources (
RX J0439.8-6809,
RX J0513.9-6951,
RX J0527.8-6954,
CAL 87,
CAL 83,
1E 0035.4-7230,
RX J0048.4-7332,
1E 0056.8-7154,
RX J0019.8 +2156,
RX J0925.7-4758,
AG Draconis)
               are
               approximated
               with
               theoretical
               spectra
               obtained
               in
               LTE
               models
               for
               the
               atmospheres
               of
               hot
white
dwarfs
with
line
blanketing.
The
confidence
intervals
of
parameters
derived
from
these
approximations $ T_{\rm eff}$,
$\log~ g$,
$N_H$, and
$R^2/d^2$ are
determined.

The
results
are
compared
with
predictions
for a
model
with
stable/recurrent
thermonuclear
burning
on
the
white-dwarf
surface.
\bigskip

\centerline {\bf 1. Introduction}
\bigskip

Supersoft
X-ray
sources
(hereafter
"supersoft
sources")
are a
class
of
X-ray
object
distinguished
by
ROSAT
observations (Tr\"umper et al. 1991, Hasinger 1994, Kahabka and van den
Heuvel, 1997).  The main characteristic of these objects is a very
soft X-ray spectrum that falls off near 0.5 - 1 keV.  Blackbody
approximations to the X-ray spectra yield temperatures from 10 to 80 eV.
As a
rule,
the
luminosities
derived
from
such
approximations
are
very
high
($\sim 10^{38}$ erg/s), and
often
exceed
the
Eddington
limit
for
solar-mass
objects.
The
first
observations
of
supersoft
sources
were
obtained
by
the
Einstein
Observatory,
but
ROSAT
was
the
first
satellite
that
was
able
to
distinguish
them
as a
distinct
class
and
detect
significant
numbers
of
these
sources.
Currently,
about
60
bright
supersoft
sources
are
known,
located
in
our
Galaxy,
the
Magellanic
Clouds,
the
Andromeda
galaxy,
and
NGC
55
(Greiner 2000).

One
widely
adopted
model
that
can
explain
the
nature
of
at
least
classical
double
supersoft
sources
is
that
of
van
den
Heuvel
et
al.
(1992),
in
which
supersoft
sources
are
close
binary
systems
containing a
white
dwarf
and a
more
massive
subgiant
secondary
that
overfills
its
Roche
lobe.
If
the
mass
of
the
latter
star
is
approximately
twice
the
mass
of
the
white
dwarf,
the
accretion
onto
the
white
dwarf
occurs
on
the
thermal
time
scale
for
the
secondary
at a
high
rate
($\sim 10^{-7} M_{\odot}/yr$),
which
gives
rise
to
stable
nuclear
burning
on
its
surface,
leading
to
the
observed
soft
X-ray
emission.
The
theoretical
possibility
of
such a
situation
was
predicted
in
(Paczynski and Zytkov 1977)
and
subsequently
studied
in
detail
by a
number
of
authors
(Iben 1982, Nomoto 1982, Fujimoto 1982, Iben and Tutukov 1996).

The
spectral
energy
distribution
of
such a
supersoft
source
should
be
described
by
the
theoretical
spectrum
of a
white-dwarf
model
atmosphere
with
the
appropriate
effective
temperature
$T_{\rm eff}$,
gravitational
acceleration
$g$,
and
chemical
composition $A$
(see
Section
3).
Atmospheres
with
$T_{\rm eff} \sim 10^5 - 10^6$ K
radiate more efficiently
at
energies
0.1 - 0.5
keV
than a
blackbody,
so
that
applying
LTE
model
atmospheres
without
taking
lines
into
account
yielded
bolometric
luminosities
for
supersoft
sources
that
were
below
the
Eddington
limit
for a
solar-mass
object
(Heise et al. 1994).

The
X-ray
spectra
of
supersoft
sources
have
been
approximated
using a
variety
of
theoretical
models,
from
blackbody
models
to
non-LTE
models
taking
line
absorption
into
account
(Hartmann et al. 1999).
The
parameters
of
most
sources
have
been
estimated
using
LTE
models
without
including
the
effect
of
line
absorption.

Our
goal
in
the
current
study
was
to
rereduce
series
of
ROSAT
observations
of
known
supersoft
sources
using a
unified
method,
in
order
to
derive
the
physical
characteristics
of
these
objects
in a
uniform
way,
approximating
the
observed
fluxes
using
theoretical
spectra
for
blanketed
(i.e.
including
the
effect
of
lines)
LTE
model
atmospheres
for
hot
white
dwarfs.
It
is
obvious
that
considering
blanketing
of
non-LTE
model
atmospheres
for
hot
white
dwarfs
would
be
more
realistic,
but
such
computations
are
appreciably
more
complex
and
time-consuming
(Hubeny and Lanz 1995).
In
addition,
we
note
that
the
temperature
structures
of
blanketed
non-LTE
and
LTE
model
atmospheres
are
very
similar
and
differ
substantially
from
non-
LTE
models
without
line
absorption
(Anderson 1990).
Therefore,
we
consider
blanketed
LTE-model
atmospheres
of
hot
white
dwarfs
to
be
more
realistic
than
non-LTE
models
without
line
absorption
(Anderson 1990).

Recent
high
resolution
($ \sim$ 0.06 \AA)
calibration
observations
of
the
source
CAL~83
by
the
XMM-
Newton
satellite
(Paerels et al 2001)
showed
that
its
spectrum
is
rich
in
absorption
lines.
This
indicates
the
photospheric
nature
of
the
spectrum
and
demonstrates
that
any
detailed
analysis
of
the
X-ray
spectra
of
these
objects
obtained
by
the
new
generation
of
space
observatories
must
make
use
of
the
method
of
synthetic
spectra.

The
full
designations
of
the
sources
are
presented
in
Table
2.
We
will
use
shortened
designations
consisting
only
of
the
first
several
symbols.

In
all,
we
studied
ten
supersoft
sources.
We
also
present
data
for
AG
Dra;
although
its
spectrum
is
very
soft,
the
resulting
parameters
are
very
uncertain
and
are
not
interpreted
further.
The
parameters
of
the
two
sources
RX~J0527
and
RX~J0513
are
the
first
derived
using
white-dwarf
model
atmospheres.
\bigskip

\centerline{\bf 2. Data reduction}
\bigskip

The
sample
of
sources
corresponds
to
the
list
of
Greiner (2000). A
number
of
objects
were
excluded
due
to
the
absence
of
satisfactory
observations
or
the
exceptional
softness
of
their
spectra,
which
led
to
very
large
uncertainties
in
their
parameters
(the
results
for
AG
Dra
demonstrate
what
occurs
in
this
case).
In
the
end,
we
used
11
sources
and
13
observations
in
our
study.

The
spectra
of
the
objects
were
extracted
from a
circular
region
with a
radius
large
enough
to
include
the
majority
of
the
photons.
Information
about
the
background
was
extracted
from a
zone
having
an
area
several
times
larger
surrounding
the
source.
To
increase
the
signal-to-noise
ratios
and
obtain a
high
correspondence
to
the
energy
resolution
of
the
detector,
the
spectral
flux
densities
were
recalculated
to
new
energy
channels
that
were
broader
than
the
instrumental
channels
(Table 1).

We
estimated
the
physical
parameters
of
the
sources
by
approximating
the
observed
spectra
with
theoretical
spectra
derived
from
computations
of
blanketed
LTE
model
atmospheres
taking
into
account
interstellar
absorption.
The
free
parameters
in
the
fitting
were
the
column
density
of
interstellar
hydrogen
$N_H$,the
effective
temperature
of
the
model
atmosphere
$T_{\rm eff}$,
and a
normalization
factor
accounting
for
the
geometrical
decrease
in
the
flux,
$R^2$/$d^2$
($R$
is
the source radius and $d$
is
the
distance
to
the
source).
The
gravitational
acceleration
$log~g$
was
fixed
at a
preliminarily
determined
optimal
value
(this
parameter
influences
the
spectrum
much
more
weakly
than
the
others;
see
Section 4).

Table 2
presents
the
main
data
on
the
sources
and
the
parameters
of
the
observations
used.
\bigskip

\centerline{\bf 3. Method for computing the hot white-dwarf model
atmospheres}
\bigskip

The model atmospheres were characterized by their effective temperature
$T_{\rm eff}$,
surface
gravitational
acceleration
$g$,
and a
scaling
factor
for
the
abundance
of
elements
heavier
than
helium
$A$.
Solar
chemical
composition
corresponds
to $A$=1,
and a
heavy-
element
content a
factor
of
ten
lower
corresponds
to $A$
=0.1.
We
assumed
that
the
atmosphere
was
stationary
and
static,
and
consisted
of
uniform,
planeparallel
layers.

The
model
atmosphere
was
found
numerically,
and
the
distribution
of
atoms
and
ions
in
terms
of
their
excitation
and
degrees
of
ionization
were
determined
by
the
Saha
and
Boltzmann
equations
assuming
local
thermodynamic
equilibrium.
We
used a
modified
version
of
the
ATLAS5
code
of
Kurucz (1970).
The
computations
included
the
15
most
widespread
elements
(H,
He,
C,
N,
O,
Ne,
Na,
Mg,
Al,
Si,
S,
Ar,
Ca,
Fe,
Ni).
In
contrast
to
the
original
code
of
Kurucz,
we
considered
all
ionization
states.
Data
on
the
opacity
cross
sections
were
taken
from
(Verner et al. 1996).
We
also
took
into
account
Thomson
scattering
on
free
electrons
(Compton
effects
in
the
atmospheres
ot
white
dwarfs
weaken
as
the
gravitational
acceleration
increases,
and
do
not
influence
the
continuum
spectrum
(Madej 1994).
We
included
about
1200
of
the
strongest
spectral
multiplets
and
lines
(for
the
ionization
states
corre-
sponding
to
the
conditions
in
the
atmosphere)
selected
from
the
list
(Verner et al. 1996),
whose
effect
was
determined
via
direct
integration.
We
took
the
line
profiles
to
be
Voigt
profiles
broadened
by
natural
damping,
the
Doppler
effect
(taking
the
microturbulent
velocity
to
be 1
km/s),
and
the
Stark
effect.
The
Stark
half-
width
was
computed
using
the
simple
approximation
formula
of
Kurucz and Furenlid (1979).
We
used a
grid
of $\sim$
8000
points
in
frequency
(from 3 $\cdot
10^{13}$
to
10$^{21}$
Hz)
and
100
points
in
depth
to
solve
the
radiative-transfer
equation
using
the
method
of
Auer (1976)
for
three
angles.
The
upper
boundary
condition
assumed
the
absence
of a
rising
flux
at
the
first
surface
point.
We
used a
diffusion approximation
as a
lower
boundary
condition.
The
temperature
structure
of
the
atmosphere
was
computed
using a
temperature-correction
method.

In
some
high-lying
points
in
the
atmosphere
(which
exert
virtually
no
influence
on
the
emergent
flux),
the
radiation
pressure
can
exceed
the
gravitational
force.
Physically,
this
implies
the
presence
of a
stellar
wind
from
the
surface
of
the
white
dwarf.
However,
the
study
of
this
wind
falls
outside
the
framework
of
the
problem
at
hand,
and
we
will
ignore
its
effect.
\bigskip

\centerline{\bf 4. Results of the computations}
\bigskip

We
computed a
grid
of
model
white-dwarf
atmospheres
with
solar
chemical
composition
(Anders and Grevesse 1989)
and
with
effective
temperatures 5 $\cdot
10^4 - 1.3 \cdot
10^6$ K in
steps
of
$10^4$ K
and
surface
gravities
$\log~g$ =7.0 -- 9.5
in
steps
of
0.5
for
comparison
with
the
observations.
We
computed
only
the
models
that
did
not
exceed
the
Eddington
limit,
i.e.,
with
$\log~g > \log~g_E$.
This
limiting
surface
gravity
$\log~g_E$
is
specified
by
the
relation
\be
      \log~g_E = 4.88 + 4\log~T5,
\ee
where
$T5 = T_{\rm eff}/10^5$
K.

The
results
of
the
computations
are
illustrated
in
Fig. 1.
Figure
1a
shows
the
temperature
structures
of
the
blanketed
and
unblanketed
models.
Figure
1b
shows
the
difference
between
the
model
spectra
without
lines
(dashed)
and
with
lines
(solid).
This
difference
is
more
clearly
visible
in
Fig.
1c,
which
shows
the
spectra
averaged
over
intervals
of
10
eV.
The
flux
of
the
blanketed
model
is
lower
at
high
energies
and
higher
at
softer
energies
than
that
of
the
unblanketed
model,
reflecting
the
differences
in
their
temperature
structures.
There
are
also
local
dips
associated
with
the
presence
of
large
numbers
of
strong
lines
in
these
regions.
Note
that
the
uncertainty
associated
with
our
poor
knowledge
of
the
magnitude
of
the
Stark
effect
for
the
spectral
lines
does
not
significantly
influence
the
resulting
spectral
energy
distribution.

Figure 2
shows
the
dependence
of
the
spectra
on
variations
of
the
model
parameters:
effective
temperature
(Fig.
2a),
surface
gravity
(Fig.
2b),
and
heavy-
element
abundance
(Fig.
2c).
At
energies
above
0.5
keV,
variations
in
the
slope
and
overall
flux
of
the
spectra
are
most
appreciable
in
the
presence
of
variations
of
the
effective
temperature.
Variations
of
$\log~g$
influence
the
overall
redistribution
of
energy
much
more
weakly,
and
variations
of
the
chemical
composition
are
manifest
only
via
differing
values
for
absorption
jumps.
\bigskip

\centerline{\bf 5. Fitting technique}
\bigskip

The
PSPC
instrument
on
ROSAT
has
the
following
characteristics
at
soft
energies.
The
energy
resolution
of
the
detector
($\Delta E/E$)
to
0.5
keV
is
about
film
shielding
the
instrument.
In
addition,
absorption
90\%.
The
detector
has
no
sensitivity
near
0.4
keV
in
the
interstellar
medium
exerts a
large
influence
at
since
photons
with
this
energy
are
absorbed
in
the
these
energies.
If
there
are
no
photons
with
energies
higher
than
0.5
keV
in
an
observed
spectrum,
variations
in
the
interstellar
absorption
may
be
compensated
by
the
joint
influence
of
the
normalization
for
the
dilution
of
the
flux
and
the
effective
temperature,
which
affects
the
slope
of
the
spectrum.
As a
result,
various
combinations
of
these
three
parameters
can
yield
statistically
similar
results.
This
is
usually
visible
in a
plot
of
confidence-level
contours
in
the
$N_H$ -- $T_{\rm eff}$
plane
(curves
bounding
zones
within
which
the
true
parameter
value
is
located
with
some
probability)
as a
characteristic
region
of
possible
$T_{\rm eff}$
and
$N_H$
values
(Fig.
3).
As a
rule,
the
upper
left
part
of
this
zone
is
consistent
with
the
Galactic
value
of
$N_H$
in
the
specified
direction
(we
will
call
this
the
Galactic
$N_H$
value)
and
yields
plausible
luminosities
for
the
source.

In
connection
with
this,
we
estimated
the
errors
in
$T_{\rm eff}$
and
the
normalization
when
limiting
the
column
density
$N_H$
in
the
region
of
most
plausible
parameter
values.
We
adopted
the
extreme
values
of
the
errors
of
$T_{\rm eff}$
and
the
normalization
for
two
fixed
values
of
$N_H$
for
the
boundary
values
delimiting
this
region.
We
first
determined
the
90\%
probability
boundaries
for
$N_H$.
Then,
we
determined
the
errors
in
the
two
other
parameters
for
two
values
of
$N_H$ -- the
minimum
value
of
$N_H$
and
either
the
maximum
value
of
$N_H$
or
double
best
fit,
depending
on
which
turned
out
to
be
closer
to
the
best
approximation.
Observations
for
which
errors
were
derived
in
this
way
are
denoted
in
Tables 3
and 4
by
the
letters A
and B
in
the
вype
of
errors
column.
Statistically,
these
errors
yield a
probability
of
about
98\%
that
the
true
value
is
within
the
error
interval.

For
several
sources,
the
parameters
were
well
constrained
in
the
zone
corresponding
to
the
best-fit
approximation.
This
is
because
the
data
for
these
sources
were
fairly
hard,
so
that
either
the
cutoff
in
their
spectra
is
beyond
0.5
keV
(RX J0513)
or
the
maximum
observed
flux
occurs
at
0.7 --- 0.9
keV
(CAL 87
and
RX J0925).
In
this
case,
we
did
not
use
the
error-estimation
method
described
above,
and
the
parameter
errors
were
calculated
using
the
usual
statistical
criteria
($\chi^2$
=2.71,
90\%
probability,
type C
in
Tables 3
and
4).

Given
the
complexity
of
accurately
estimating
parameters,
we
attempted
to
determine
the
most
general
bounds
for
the
possible
parameter
values.
Most
often,
due
to
the
impossibility
of
separating
the
influences
of
individual
parameters,
the
magnitude
of
$\chi^2$
remained
virtually
unchanged
for
any
value
of
the
surface
gravity;
however,
different
effective
temperatures
were
naturally
obtained
for
different
surface
gravities.
Therefore,
we
present
two
approximations
for
nearly
all
the
sources,
corresponding
to
the
lower
and
upper
bounding
values
of
the
surface
gravity.

In
addition,
very
often,
very
broad
limits
for
the
column
density
$N_H$
were
allowed
statistically.
Our
analysis
(Section
6)
showed
that
the
Galactic
column
density
(Dickey and Lockman 1990)
is
located
within
the
confidence
intervals
for
most
of
the
sources
(located
in
the
Large
and
Small
Magellanic
Clouds).
Accordingly,
we
have
included
approximations
for
the
source
parameters
obtained
by
fixing
the
column
density
to
be
the
Galactic
value
(the
errors
were
calculated
in
the
usual
way,
error
type
C).
\bigskip

\centerline{\bf 6. Discussion}
\bigskip

The
approximation
results
are
shown
in
Tables 3
and
4,
which
contain
the
approximation
data
for
$\log~g$
=7.5
and
9.5,
and
Tables 5
and
6,
which
contain
the
data
for
these
surface
gravities
and
the
Galactic
value
of
$N_H$.
Figure 3
depicts
the
spectra
and
confidence
contours
for
the
surface
gravities
used
in
the
analyses
for
two
sources.

Figure 4
compares
our
results
with
the
data
of
other
authors.
Figure
4a
shows
such a
comparison
for
the
effective
temperatures.
Our
data
for
the
minimum
$\log~g$
are
shown
for
all
the
sources;
for
RX
J0439,
RX
J0527,
CAL
83,
and
1E
0035,
we
have
used
the
approximation
with
the
Galactic
$N_H$
value,
while
$N_H$
has
been
left
as a
free
parameter
for
the
other
sources.
Overall,
the
agreement
is
good,
though
we
should
point
out a
number
of
features.

First,
the
temperatures
for
RX
J0527
and
RX
J0513
are
approximately
100 000 K
higher
than
those
determined
by
other
authors
based
on
approximations
of
their
spectra
using
Planck
functions.
Second,
for
the
limiting
values
of
$T_{\rm eff}$
for
the
two
hardest
sources
CAL
87
and
RX
J0925,
we
used
temperatures
derived
from
non-LTE
models
(lower
bounds)
and
unblanketed
LTE
models
(upper
bounds)
(Hartmann et al 1999, Hartmann and Heise 1997).
We
can
see
that
the
non-LTE
values
are
lower
and
the
LTE
values
(for
the
non
blanketed
models)
higher
than
our
values.
The
remaining
sources
approximated
by
non
blanketed
LTE
models
also
display
somewhat
higher
temperatures
than
those
we
derived.
Consequently,
we
conclude
that
including
the
effect
of
lines
in
the
LTE
models
for
the
white-dwarf
atmospheres
leads
to
lower
temperatures
for
the
supersoft
sources.

Figure
4b
compares
our
values
for
$N_H$
in
the
directions
of
the
sources
with
previously
published
values.
We
adopted
the
extreme
values
from
the
literature
data
as
the
limiting
values
of
$N_H$. If
they were known,
we
used
independent
values
of
$N_H$
derived
from
optical
and
ultraviolet
observations.
The
figure
shows
our
resulting
values
for
the
minimum
$\log~g$.

We
used
the
data
of
the
Greiner's catalog (2000),
as
well
as
from
the
following
papers:
1E
0056 - (Kahabka et al. 1994)
($N_H$),
RX
J0048 - (Kahabka et al. 1994),
1E
0035 - (Kahabka et al. 1999),
RX
J0019 - (Beuermann et al 1996),
RX
J0925 - (Motch et al. 1994)
($N_H$).

The
gravitational
acceleration
is
poorly
constrained.
We
were
able
to
fix
it
well
only
for
the
three
sources
with
the
highest
temperatures:
RX
J0925,
CAL 87
(the only possible value
for them
is
log g =
9.5),
and
RX
J0513
(log g
=8.4+0.04,
errors
shown
are).
The
errors
in
their
$\log~g$
values
in
Fig. 5
were
estimated
to
be
$\pm$ 0.5.
For
the
remaining
sources,
values
of
$\log~g$
from
the
entire
range
of
parameters
for
our
model
grid
are
statistically
allowed.
However,
the
model
of
van
den
Heuvel
et
al. (1992)
predicts a
rigorous
relation
between
$T_{\rm eff}$
and
$\log~g$,
according
to
which
stable
hydrogen
burning
is
possible
without
an
appreciable
increase
in
the
white-dwarf
radius
only
within a
rather
narrow
range
of
accretion
rates,
called
the
stable-burning
strip
(SBS)
(Nomoto 1982, Fujimoto 1982, Iben and Tutukov 1996).
The
edges
of
the
SBS
depend
on
the
mass
of
the
white
dwarf.
When
the
accretion
rate
$\dot M$
is
lower,
matter
will
accumulate
on
the
surface
of
the
white
dwarf,
with
subsequent
explosive
burning
and
the
ejection
of
the
accreting
envelope
(novae,
recurrent
novae,
and
symbiotic
novae).
At
higher
values
of
$\dot M$, there
will
be
continuous
stable
hydrogen
burning
with a
luminosity
close
to
the
Eddington
luminosity.
The
excess
matter
will
be
blown
out
by
the
optically
thick
wind
from
the
white-dwarf
surface,
leading
to
an
increase
in
the
effective
radius
of
the
photosphere
and a
decrease
in
the
effective
temperature
(Kato 1996).
Using
the
theoretical
relation
between
the
mass
and
radius
of a
white
dwarf
(Popham and Narayan 1995),
we
can
reflect
the
SBS
from
the
$\dot M$ --- M
(Fujimoto 1982)
to
the
$T_{\rm eff}$ ---
$\log~g$
(van Teeseling et al. 1996)
plane.
The
SBS,
Eddington
limit,
and
positions
of
the
sources
in
these
coordinates
are
shown
in
Fig.
3.
The
three
sources
with
well-defined
values
of
$\log~g$
lie
inside
the
SBS
(within
the
errors
in
the
parameters).
The
temperatures
of
the
remaining
sources
are
also
consistent
with
the
SBS
for
the
case
of
the
minimum
allowed
surface
gravity
($\log~g$ = 7.5 - 8).

The
computations
of
models
for
white
dwarfs
with
stable
surface
thermonuclear
burning
predict
an
increase
in
the
photospheric
radius
in
the
SBS
by a
factor
of
two
to
three
compared
with
the
radius
of a
cool
white
dwarf
(Iben 1982, Fujimoto 1982).
However,
using
source
sizes
exceeding
the
radius
of a
cool
white
dwarf
of
the
given
mass
by a
factor
of
two
when
reflecting
the
stable-
burning
strip
onto
the
$T_{\rm eff}$ ---
$\log~g$
plane only leads to a
shift
of
the
strip
downward
to
the
left
along
the
strip
itself
and
the
Eddington
limit.
Therefore,
in
this
case,
the
conclusions
drawn
above
that
the
sources
are
located
in
the
stable-burning
strip
remain
valid.
\bigskip

\centerline{\bf 7. Conclusion}
\bigskip

We
have
carried
out
an
analysis
of
archival
ROSAT
observations
of
11
known
supersoft
X-ray
sources.
We
have
derived
the
atmospheric
parameters
$T_{\rm eff}$
and
$\log~g$
for
these
sources
by
approximating
their
spectra
using
computed
theoretical
spectra
for
blanketed
LTE
models
of
hot
white-dwarf
atmospheres.
The
resulting
parameter
values
are
in
agreement
with
values
published
previously.

In
the next our
paper (Suleimanov and Ibragimov 2003),
we
will
carry
out
an
analysis
of
the
parameters
obtained
from
the
point
of
view
of
their
consistency
with
the
theory
of
stable/recurrent
burning;
luminosity,
mass,
and
radius
estimates
for
seven
sources;
and a
discussion
of
the
effective
temperature - mass
relation.
\bigskip

\centerline{\bf 8. Acknowledgments}
\bigskip

This
work
was
supported
by
the
Russian
Foundation
for
Basic
Research
(project
N
99-02-17488
and
02-02-17174).
A.
A.
Ibragimov
thanks
the
High-
energy
Astrophysics
Department
of
the
Space
Research
Institute
of
the
Russian
Academy
of
Sciences
for
hospitality
during
the
time
the
work
was
being
carried
out.
\bigskip

\centerline{\bf References}
\bigskip

E.
Anders
and
N.
Grevesse,
Geochim.
Cosmochim.
Acta 1989,
53,
197

L.
Anderson,
Astron.
Soc.
Pac.
Conf.
Ser. 1990,
77

L.
Auer,
J.
Quant.
Spectrosc.
Radiat.
Transf. 1976,
16,
931

K.
Beuermann,
K.
Reinsch,
H.
Barwig,
et
al., Astron.
Astrophys. 1995,
294,
L1

J.
Dickey
and
F.
Lockman,
Ann.
Rev.
Astron.
Astrophys. 1990,
28,
215

J.
Greiner,
New
Astron. 2000,
5,
137

M. Fujimoto,
Astrophys. J. 1982,
257,
767

H.
Hartmann
and
J.
Heise,
Astron.
Astrophys. 1997,
322,
591

H.
Hartmann,
J.
Heise,
P.
Kahabka,
C.
Motch,
and
A.
Parmar,
Astron.
Astrophys. 1999,
346,
125

G.
Hasinger,
in
Reviews
in
Modern
Astronomy,
Ed.
by
G.
Klare
1994,
7,
129

J.
Heise,
A.
van
Teeseling,
and
P.
Kahabka,
Astron.
Astrophys. 1994,
288,
L45

I.
Hubeny
and
T.
Lanz,
Astrophys. J. 1995,
439,
875

I.
Iben,
Astrophys. J. 1982,
259,
244

I.
Iben
and
A.
V.
Tutukov,
Astrophys.J.
Suppl.
Ser. 1996,
105,
145

P.
Kahabka,
W.
Pietsch,
and
G.
Hasinger,
Astron.
Astrophys. 1994,
288,
538

P.
Kahabka
and
van
den
E.
P.
J.
Heuvel,
Ann.
Rev.
Astron.
Astrophys. 1997,
35,
69

P.
Kahabka,
A.
Parmar,
and
H.
Hurtmann,
Astron.
Astrophys. 1999,
346,
453

M.
Kato,
in
Supersoft
X-Ray
Sources,
Ed.
by
J.
Greiner;
Lect.
Notes
Phys. 1996,
472,
15

R.
Kurucz,
Smithson.
Astrophys.
Obs.
Spec.
Rep. 1970,
309, 1

R.
Kurucz
and
I.
Furenlid,
Smithson.
Astrophys.
Obs.
Spec.
Rep. 1979,
387, 1

J.
Madej,
Astron.
Astrophys. 1994,
286,
515

C.
Motch,
G.
Hasinger,
and
W.
Pietsch,
Astron.
Astrophys. 1994,
284,
827

K.
Nomoto,
Astrophys. J. 1982,
253,
798

B.
Paczynski
and
A.
Zytkow,
Astrophys. J. 1978,
222,
604

F.
Paerels,
A.
Rasmussen,
H.
Hurtmann,
et
al.,astro-
ph/0011038
(2000).

R.
Popham
and
R.
Narayan,
Astrophys. J. 1995,
442,
337

V.
F.
Suleimanov
and
A.
A.
Ibragimov,
Astron.
Rep. 2003,
47,
197

J. Tr\"umper, G. Hasinger, and B.
Aschenbach,
et
al.,
Nature, 1991,
349,
579

E.
P.
J.
van
den
Heuvel,
D.
Bhattacharya,
K.
Nomoto,
and
S.
Rappaport,
Astron.
Astrophys. 1992,
262,
97

A.
van
Teeseling,
J.
Heise,
and
P.
Kahabka,
in
IAU
Symp.
165:
Compact
Stars
in
Binaries,Ed.
by
J.
van
Paradijs,
E.
P.
J.
van
den
Heuvel,
E.
Kuulkers,
et
al.
(Kluwer
Academic,
Dordrecht,
1996),
p.
445.

D.
A.
Verner,
G.H.Ferland,K.T.Korista,and
D.
G.
Yakovlev,
Astrophys. J. 1996,
465,
487

D.
A.
Verner,
E.
M.
Verner,
and
L.
J.
Ferland,
Bull.
Am.
Astron.
Soc. 1996,
188,
54.18

\newpage
\begin{table}
\caption{Grouping of counts in broad channels (from one-third to one-half
the energy resolution of the detector) used in the analysis.}
\begin{center}
\begin{tabular}{|c|c||c|c|}
\hline
Channel & Energy  & Channel & Energy \\
number &  band, keV & number &  band, keV \\
\hline
\hline
1 & 0.16 -- 0.20  & 8 & 0.87 -- 1.04 \\
2 & 0.21 -- 0.26  & 9 & 1.05 -- 1.24 \\
3 & 0.27 -- 0.34  & 10 & 1.25 -- 1.46 \\
4 & 0.35 -- 0.44  & 11 & 1.47 -- 1.69 \\
5 & 0.45 -- 0.56  & 12 & 1.70 -- 1.93 \\
6 & 0.57 -- 0.70  & 13 & 1.94 -- 2.23 \\
7 & 0.71 -- 0.86  & & \\
\hline
\end{tabular}
\end{center}
\end{table}

\newpage
\begin{table}[htbp]
\caption{
Data
on
sources
and
observations
used}
\begin{center}
\begin{tabular}{llllll}
\hline
Object & Observation & $\alpha$ & $\delta$   & Count & Off-axis \\
       &            &          &             & rate, phot/s   &  angle \\
\hline
\hline
RX J0439.8-6809 & rp400161n00 & $04^h39^m49_.^s6$ & $-68^o09'01''$ &1.40 & 1.$'$52 \\
RX J0513.9-6951 & rp900398a02 & $05^h13^m48_.^s8$&$-69^o52'00''$ &1.947 & 0.57 \\
RX J0527.8-6954 & rp400148n00 & $05^h27^m48_.^s6$& $-69^o54'02''$   & 0.1171 & 0.061\\
CAL 87          &rp400012n00 & $05^h46^m45_.^s0$& $-71^o08'54''$  &0.121 & 0.1 \\
CAL 83 & rp110180n00 & $05^h43^m33_.^s5$&$-68^o22'23''$    &0.53 &40.29 \\
\hline
1E 0035.4-7230  & rp400299n00 & $00^h37^m19_.^s0$&$-72^o14'14''$ &0.52 &0.167
\\
& rp400149n00 & & &0.41 &0.167 \\
RX J0048.4-7332 & rp600196a01 &$00^h48^m20_.^s0$ &$-73^o31'55''$ &0.17 & 20.87\\
1E 0056.8-7154  & rp600455a02 & $00^h58^m37_.^s1$&$-71^o35'48''$ & 0.325 & 18.3\\
&rp400300a01 & & &0.373 & 0.38\\
\hline
RX J0019.8+2156 & rp400322n00 &$00^h19^m50_.^s1$ &$+21^o56'54''$ &1.96 &0.103\\
RX J0925.7-4758 & rp900377n00 & $09^h25^m42.^s0$ & $-47^o58^m00^s$ & 0.675 &40.99 \\
AG Draconis     & rp200689n00 &$16^h01^m41.^s1$ &$+66^o48'10''$ & 1.09 &1.88 \\ \hline
\end{tabular}
\end{center}
\label{table2}
\end{table}

\newpage
\renewcommand{\arraystretch}{1.5}
\begin{table}[htbp]
\caption{
Approximation
parameters
for
the
minimum
$\log~g$.}\label{data1}
\begin{center}
\small
\begin{tabular}{l|c|c|c|c|c|c|c}
\hline
Object   & $N_{{H}}$,     & $T_{{eff}},$  & $log~g$ &
$log~(R/d)^2$      & Flux &  $\chi^2/$ d.o.f. & Type of \\
                    &$10^{20}$ cm$^{-2}$ & $\times 10^5$ K & cm s$^{-2}$&
& (0.2-2 keV) & & errors \\  & & & & & erg cm$^{-2}$ s$^{-1}$ & & \\
\hline
\hline
RX J0439 &
$4.75_{-4.67}^{+10.15} $ & $2.80_{-0.70}^{+1.52}$ & 7.5 &
$-26.63_{-2.84}^{+5.29}$ & $4.32\cdot 10^{-11}$ & 9.30/8
& A \\
RX J0513 &
$5.94_{-0.40}^{+0.47}  $ & $5.95_{-0.07}^{+0.10}$ & $8.4_{-0.15}^{+0.04}$ &
$-28.57_{-0.08}^{+0.10}$ & $1.09\cdot 10^{-10}$ & 19.4/8 & C \\
RX J0527 &
$27.6_{-24.25}^{+11.3} $ & $3.02_{-1.81}^{+2.49}$ & 8.0 &
$-23.43_{-6.65}^{+10.0}$ & $1.26\cdot 10^{-7} $ & 1.47/8 & A\\
CAL 87          &
 $69.2_{-22.0}^{+23.1}  $ & $8.20_{-0.24}^{+0.22}$ & 9.0 &
$-28.20_{-0.81}^{+0.85}$ & $1.17\cdot 10^{-9} $ & 13.5/8  & C \\
CAL 83  & $11.0_{-6.46}^{+11.3}$ & $4.64_{-0.64}^{+0.94}$ & 8.0
& $-27.59_{-1.51}^{+2.17}$ & $2.62\cdot 10^{-10}$&  16.1/8 & B \\
1E 0035  & $5.00_{-2.34}^{+3.72}$ & $4.82_{-0.29}^{+0.49}$ & 8.0
& $-28.89_{-0.63}^{+0.80}$ &  $1.67\cdot 10^{-11}$& 6.69/8 & C\\
& $3.83_{-1.89}^{+3.13}  $ & $4.96_{-0.34}^{+0.54}$ & 8.0  &
$-29.29_{-0.56}^{+0.74}$ & $ 8.16 \cdot 10^{-12}$ & 3.42/8& C \\
RX J0048
&  $27.6_{-12.6}^{+10.5}  $ & $3.35_{-0.48}^{+0.55}$ & 7.5 &
$-24.55_{-2.50}^{+2.82}$ & $3.09 \cdot 10^{-8} $ & 8.08/8 & C\\
1E 0056  &
 $12.1_{-9.24}^{+7.00}  $ & $2.69_{-0.43}^{+0.95}$ & 7.5 &
$-25.01_{-3.82}^{+1.37}$ & $1.20 \cdot 10^{-9}$ & 9.41/8 & A\\
& $3.71_{-2.38}^{+21.29}$ & $4.00_{-0.97}^{+1.24}$ & 8.0
& $-28.85_{-0.44}^{+2.02}$ &  $5.31\cdot 10^{-12}$& 13.2/8& B\\
RX J0019 &  $16.7_{-13.35}^{+6.9} $ & $2.80_{-0.35}^{+1.26} $ &
7.5 & $-23.62_{-4.67}^{+2.82}$ & $ 4.46\cdot 10^{-8}$&18.1/8  & A\\
RX J0925 &  $163_{-46}^{+14}      $ & $9.85_{-0.37}^{+1.05} $ &
9.5 & $-26.23_{-1.57}^{+0.59}$ &$2.53 \cdot 10^{-7}$ & 8.85/8 & C\\
AG Dra & $0.0595_{-0.0595}^{+0.0405} $ &$ 3.09_{-2.59}^{+0.41}$ &7.5
&$-28.82_{-0.18}^{+14.82}$  &  $7.67\cdot 10^{-13}$&4.21/8 & B\\
\hline
\multicolumn{8}{l}{
Note:
Fluxes
have
been
corrected
for
interstellar
absorption.}\\
\multicolumn{8}{l}{
Type
of
errors: A
denotes
error
limits
based
on
the
limiting
values
of
$N_H$
(98\%);}\\
\multicolumn{8}{l}{
B
denotes
error
limits
based
on
the
left
boundary
of
$N_H$
and
2$N_H$
for
the
best
approximation
(98\%);}\\
\multicolumn{8}{l}{
C
denotes
errors
derived
from
usual
$\chi^2$
criteria
(90\%).
The
error
for
$\log~g$
for
RXJ0513
is
1$\sigma$;}\\
\multicolumn{8}{l}{
the
remaining
parameters
were
derived
with
this
value
fixed.
The
data
for
AG
Dra
are
approximate.}\\
\end{tabular}
\end{center}
\end{table}

\newpage
\begin{table}[htbp]
\caption{
Approximation
parameters
for
the
maximum
$\log~g$.}
\begin{center}
\small
\begin{tabular}{l|c|c|c|c|c|c|c}
\hline
Object & $N_{{H}}$,     & $T_{{eff}},$  & $log~g$ &
$log~(R/d)^2$      & Flux &  $\chi^2/$ d.o.f. & Type of\\
            &$10^{20}$ cm$^{-2}$ & $\times 10^5$ K & cm s$^{-2}$&
& (0.2-2 keV) & & errors\\ & & & & & erg cm$^{-2}$ s$^{-1}$ & & \\
\hline
\hline
RX J0439 & $ 5.00_{-5.00}^{+0.06} $& $3.10_{-0.88}^{+2.75}$ & 9.5
& $-26.79_{-3.31}^{+1.81}$ & $5.03\cdot 10^{-11}$& 9.31/8 & A \\
RX J0513 &  \multicolumn{6}{c}{Not presented: one posible
value of $\log g$}\\
RX J0527 & $21.9_{-18.55}^{+8.00}$ & $3.43_{-0.22}^{+3.62}$ & 9.5
& $-25.10_{-5.52}^{+2.03}$ & $6.72\cdot 10^{-9}$& 1.47/8 & A\\
CAL 87          & \multicolumn{6}{c}
{Not presented: one possible value of $\log g$} \\
CAL 83
&  $10.6_{-6.25}^{+24.4}  $ & $5.65_{-1.08}^{+1.48}$ & 9.5 &
$-28.11_{-1.51}^{+2.19}$ &$2.16\cdot10^{-10} $&  16.3/8 & B \\
1E 0035  & $4.62_{-2.21}^{+3.54}  $ & $5.93_{-0.50}^{+0.75}$ & 9.5
& $-29.4_{-0.65}^{+0.85}$  & $ 1.44 \cdot 10^{-11}$ & 6.61/8  & C\\
&$3.51_{-1.77}^{+2.97}$ & $6.14_{-0.53}^{+0.82}$ & 9.5
& $-29.79_{-0.56}^{+0.72}$ &  $7.23\cdot 10^{-12}$&3.44/8& C \\
RX J0048  & $22.7_{-9.2}^{+5.1} $& $4.10_{-0.63}^{+0.90} $& 9.5
& $-25.88_{-0.81}^{+1.71}$ &   $5.21\cdot 10^{-9}$&9.35/8 & C\\
1E 0056   & $10.1_{-8.07}^{+4.10} $& $3.00_{-0.56}^{+1.96}$ & 9.5
& $-25.81_{-3.96}^{+2.46}$ &  $3.35\cdot 10^{-10}$& 9.82/8 & A\\
& $3.28_{-2.21}^{+17.42} $ & $4.59_{-1.24}^{+1.97}$ & 9.5  &
$-29.29_{-1.20}^{+1.95}$ &$4.26\cdot 10^{-12}$ & 13.1/8& B\\
RX J0019 & $11.7_{-8.76}^{+6.20} $& $3.31_{-0.60}^{+2.04}$ & 9.5
& $-25.40_{3.57}^{+2.81}$ &  $2.39\cdot 10^{-9}$&18.2/8  & A\\
RX J0925 & \multicolumn{6}{c}{Not presented: one possible
value of $\log g$}\\
AG Dra &  $2.51_{-2.51}^{+3.99} $ & $1.16_{-0.66}^{+2.84}$ &
9.5 & $-20.96_{-8.36}^{+4.66}$ & $3.49 \cdot 10^{-12}$ & 4.20/8 & A\\
\hline
\multicolumn{8}{l}{
Note: See comments to Table 3.}\\
\end{tabular}
\end{center}
\end{table}

\newpage
\begin{table}[htbp]
\caption{
Approximation
parameters
for
the
minimum
$\log~g$
and
Galactic
$N_H$.}\label{data2}
\begin{center}
\small
\begin{tabular}{l|c|c|c|c|c|c}
\hline
Object & $N_{{H}}$,     & $T_{{eff}},$  & $log~g$ &
$log~(R/d)^2$      & Flux &  $\chi^2/$ d.o.f. \\
             &$10^{20}$ cm$^{-2}$ & $\times 10^5$ K & cm s$^{-2}$&
& (0.2-2 keV) & \\  & & & & & erg cm$^{-2}$ s$^{-1}$ &  \\
\hline
\hline
RX J0439 & 5.60 &$ 2.72_{-0.15}^{+0.12}$& 7.5 & $-26.24_{-0.26}^{+0.35}$
&$7.99\cdot 10^{-11}$&  9.30/9 \\
RX J0513 &  7.24 &$ 5.74_{-0.03}^{+0.03}$& 8.4 & $-28.29_{-0.01}^{+0.02}$
&$1.71\cdot 10^{-10}$&  33.8/9 \\
RX J0527 &  6.31 &$ 4.65_{-1.12}^{+0.24}$& 8.0 & $-29.18_{-0.09}^{+0.83}$
&$6.90\cdot 10^{-12}$&  2.17/9 \\
CAL 87      & 7.58 & \multicolumn{4}{c}{
Unsatisfactory approximation} \\
CAL 83   & 6.33 & $5.04_{-0.17}^{+0.17}$ & 8.0 & $-28.57_{-0.08}^{+0.08}$
& $4.59\cdot 10^{-11}$&17.3/9 \\
1E 0035  & 6.94 & $4.64_{-0.11}^{+0.09}$ & 8.0 & $-28.45_{-0.06}^{+0.06}$ &
$3.62\cdot 10^{-11}$&7.58/9 \\
 & 6.94 &$ 4.64_{-0.13}^{+0.09} $& 8.0 & $-28.56_{-0.05}^{+0.08}$
&$2.83\cdot 10^{-11}$& 6.10/9 \\
RX J0048  & 4.24 & \multicolumn{4}{c}{
Unsatisfactory approximation} \\
1E 0056  & 6.16 &$ 2.90_{-0.11}^{+0.09} $& 7.5 & $-27.02_{-0.22}^{+0.26}$
&$2.56\cdot 10^{-11}$& 10.3/9\\
& 6.16 & $3.20_{-0.09}^{+0.99?}$ & 8.0 & $-27.63_{-0.86}^{+0.36}$
&$1.85\cdot 10^{-11}$& 13.3/9\\
RX J0019  & 4.20 & $3.57_{-0.20}^{+0.24} $& 7.5 & $-27.71_{-0.22}^{+0.20}$
&$3.61\cdot 10^{-11}$&19.5/9\\
RX J0925 & 55.6 &\multicolumn{4}{c}{
Unsatisfactory approximation} \\
AG Dra &  3.08 & $2.35_{-0.16}^{+0.17}$ & 7.5 &$-26.41_{-0.41}^{+0.46}$
& $7.79\cdot 10^{-12}$&4.73/9 \\
\hline
\multicolumn{7}{l}{
Note: See comments to Table 3.}\\
\end{tabular}
\end{center}
\end{table}

\newpage
\begin{table}[htbp]
\caption{
Approximation
parameters
for
the
maximum
$\log~g$
and
Galactic
$N_H$.}
\begin{center}
\small
\begin{tabular}{l|c|c|c|c|c|c}
\hline
Object &  $N_{{H}}$,     & $T_{{eff}},$  & $log~g$ &
$log~(R/d)^2$      & Flux &  $\chi^2/$ d.o.f. \\
               &$10^{20}$ cm$^{-2}$ & $\times 10^5$ K & cm s$^{-2}$&
& (0.2-2 keV) & \\  & & & & & erg cm$^{-2}$ s$^{-1}$ & \\
\hline
\hline
RX J0439 & 5.60 & $3.02_{-0.20}^{+0.19}$ & 9.5 & $-26.47_{-2.35}^{+0.37}$
&$7.77\cdot 10^{-11}$& 9.31/9 \\
RX J0513 &  7.24 & \multicolumn{4}{c}{
Not presented: one possible value of $\log g$} \\
RX J0527  & 6.31 & $5.60_{-1.68}^{+0.37}$ & 9.5 & $-29.59_{-0.14}^{+0.95}$
& $6.84\cdot 10^{-12}$&2.01/9 \\
CAL 87   & 7.58 & \multicolumn{4}{c}
{Not presented: one possible value of $\log g$} \\
CAL 83     & 6.33 &$ 6.20_{-0.25}^{+0.24}$& 9.5 & $-29.01_{-0.08}^{+0.09}$
&$4.59\cdot 10^{-11}$&  17.1/9 \\
1E 0035  & 6.94 &$ 5.61_{-0.27}^{+0.14} $& 9.5 & $-30.88_{-0.06}^{+0.11}$
&$3.59\cdot 10^{-11}$&  7.95/9 \\
 & 6.94 & $5.60_{-0.30}^{+0.14}$ & 9.5 & $-28.98_{-0.06}^{+0.13}$ &
$2.81\cdot 10^{-11}$&6.83/9 \\
RX J0048  & 4.24 & \multicolumn{4}{c}{
Unsatisfactory approximation} \\
1E 0056  & 6.16 & $3.31_{-0.20}^{+0.16}$ & 9.5 & $-27.37_{-0.29}^{+0.35}$
& $2.50\cdot 10^{-11}$&10.3/9\\
 & 6.16 &$ 3.65_{-0.26}^{+0.87} $& 9.5 & $-27.91_{-0.77}^{+0.41}$
&$1.81\cdot 10^{-11}$& 13.3/9\\
RX J0019  & 4.20 & $4.06_{-0.17}^{+0.29}$ & 9.5 & $-28.01_{-0.21}^{+0.14}$
& $3.47\cdot 10^{-11}$&19.3/9\\
RX J0925  &      & \multicolumn{4}{c}{
Not presented: one possible value of $\log g$} \\
AG Dra & 3.08 & $1.44_{-0.23}^{+0.46} $& 9.5 &$-22.49_{-2.04}^{+1.46}$
&$5.24\cdot 10^{-12}$&4.22/9 \\
\hline
\multicolumn{7}{l}{
Note: See comments to Table 3.}\\
\end{tabular}
\end{center}
\end{table}

\newpage
\centerline{\bf Captions to figures}
\bigskip

{\bf Figure 1.}
Difference
between
blanketed
and
unblanketed
white-dwarf
model
atmospheres:
(a)
temperature
structure
($m$
is
the
column
density),
(b)
spectra,
and
(c)
spectra
averaged
over
10
eV.
The
model
parameters
are
$T_{\rm eff}$ =
500 000
K,
$\log~g$ =8.5, $A$=1.
The
solid
curve
shows
the
model
with
lines,
while
the
dotted
and
dashed
curves
show
the
model
without
lines.
\bigskip

{\bf Figure 2.}
Atmospheric
model
spectra
as
functions
of
the
(a)
effective
temperature
($\log~g$=8.5,
solar
chemical
composition,
$T_{\rm eff}
=(1-9) \cdot
10^5$
K),
(b)
surface
gravity
($T_{\rm eff}=5 \cdot 10^5$ K,
solar
chemical
composition,
$\log~g$=7.5 - 9.5),
and
(c)
chemical
composition
($T_{\rm eff}= 5 \cdot 10^5$ K,
$\log~g$=8.5,
chemical
compositions
of
0.25,
0.5,and 1
of
the
solar
value).
The
spectra
are
averaged
over
intervals
of
10
eV.
\bigskip

{\bf Figure 3.}
Observed
X-ray
spectra
for
two
of
the
studied
sources
together
with
the
best-fitting
theoretical
spectra
and
the
regions
of
admissible
parameters
in
the
$N_H$ --- $T_{\rm eff}$
plane
corresponding
to
the
surface
gravities
chosen
for
the
analysis
(see
Section
6).
Values
of
the
residuals
are
presented
together
with
the
spectra.
The
contours
bound
the
68,
90,
and
99\%
probability
zones,
and
the
crosses
mark
the
best-fit
values.
The
source
CAL
87
has a
relatively
hard
spectrum,
and
its
parameters
are
localized
with
certainty;
CAL
83
has a
soft
spectrum,
and
the
zone
of
allowed
parameters
in
the
$N_H$ --- $T_{\rm eff}$
plane
is
very
extended.
\bigskip

{\bf Figure 4.}
Comparison
of
derived
parameters
with
the
data
of
other
studies
(references
given
in
the
text):
(a)
effective
temperatures
and
(b)
column
densities
of
interstellar
hydrogen.
\bigskip

{\bf Figure 5.}
Position
of
the
studied
sources
in
the
$T_{\rm eff}$ ---
$\log~g$
plane.
The
bold
solid
curves
show
the
strip
of
stable
burning.
The
dotted
curve
indicates
the
Eddington
limit.

\begin{figure}
\includegraphics[width=\columnwidth, bb=14 50 581 828, clip]{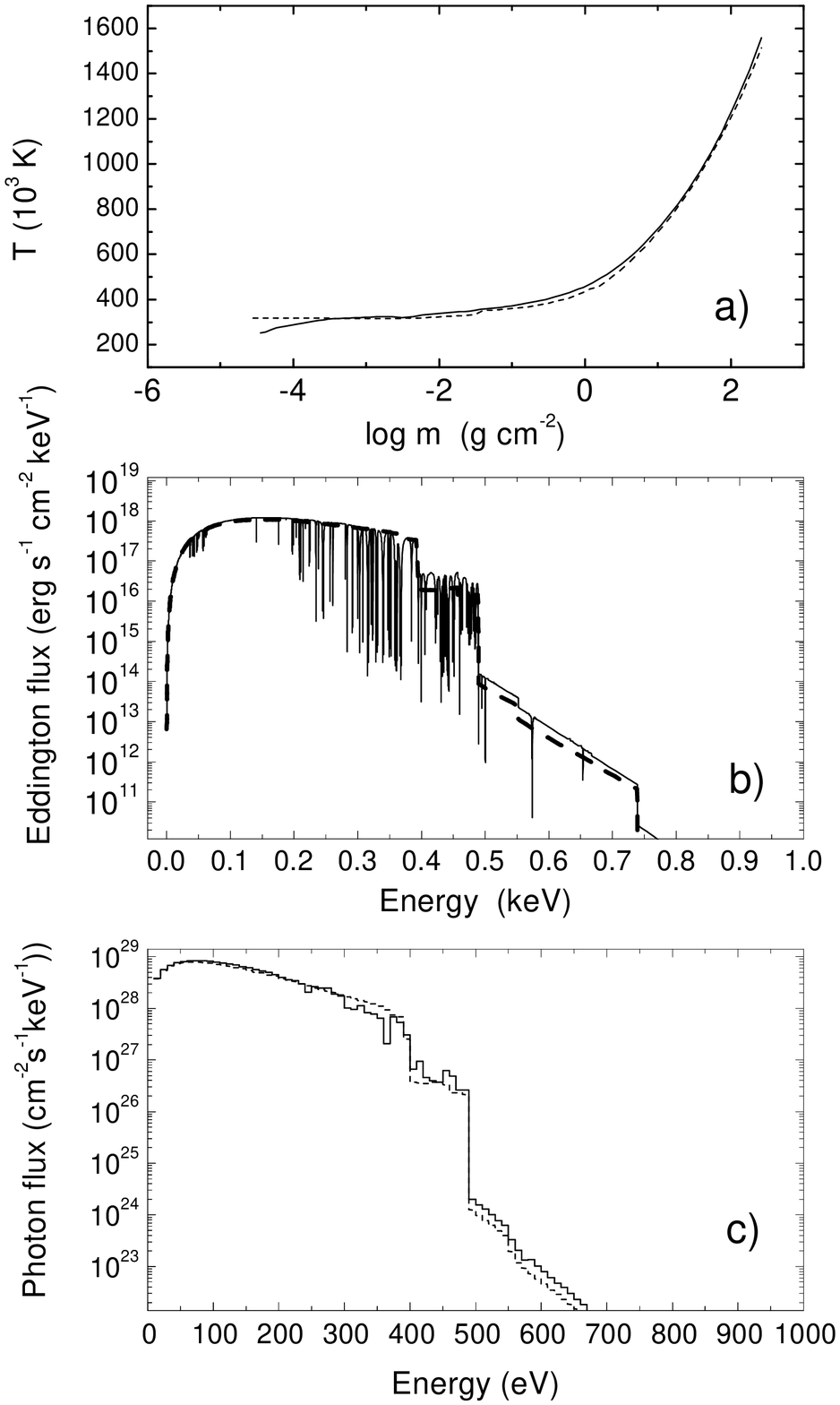}
\caption{\label{fig1}
}
\end{figure}
\begin{figure}
\includegraphics[width=\columnwidth, bb=14 50 581 828, clip]{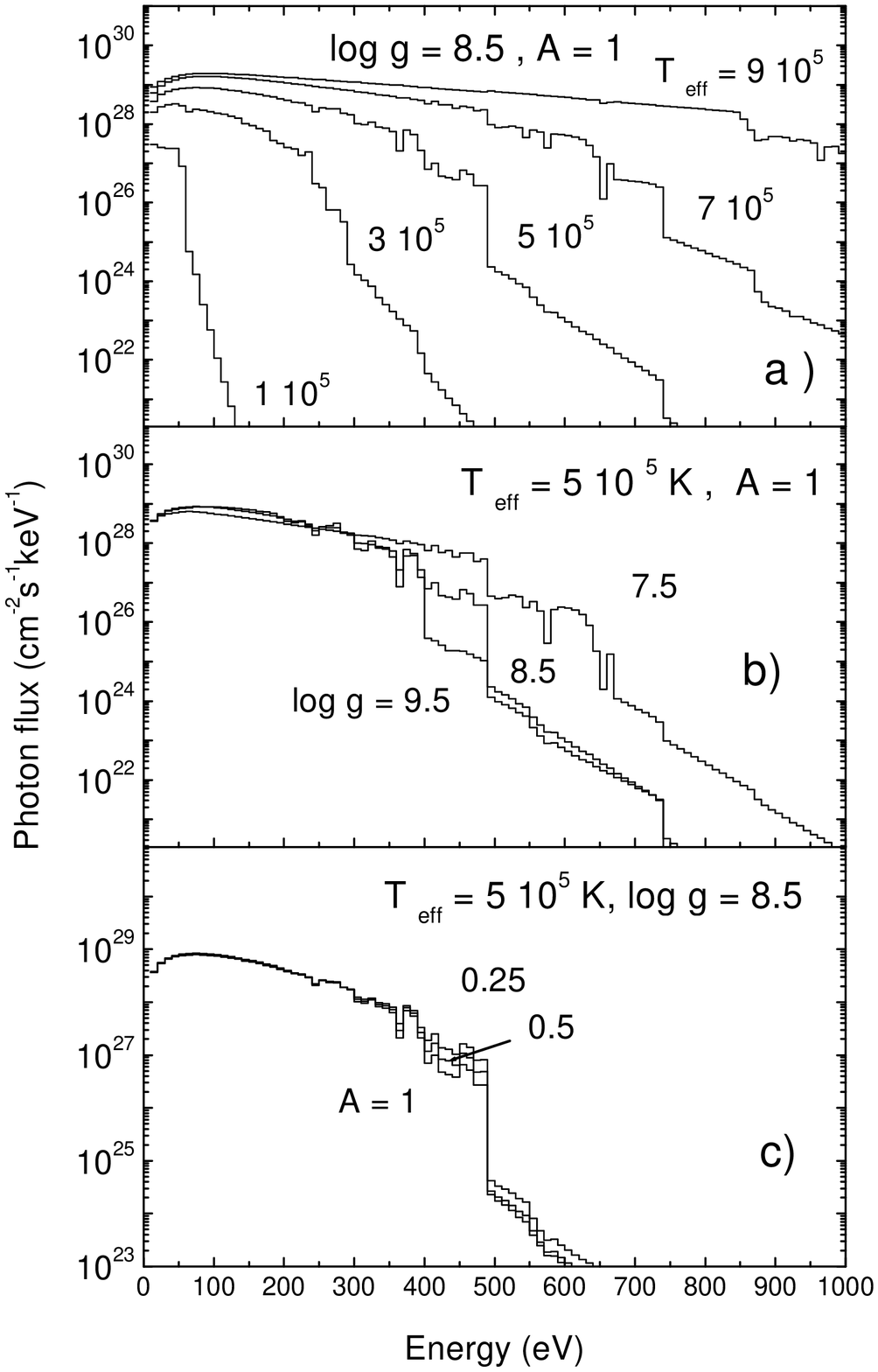}
\caption{\label{fig2}
}
\end{figure}
\begin{figure*}[htb]
\hbox{
\includegraphics[width=0.45\columnwidth,bb=20 10 600 716,clip]{Cal87l.ps}
\includegraphics[width=0.45\columnwidth,bb=20 10 600 716,clip]{Cal87c.ps}
}
\hbox{
\includegraphics[width=0.45\columnwidth,bb=20 10 600 716,clip]{Cal83l.ps}
\includegraphics[width=0.45\columnwidth,bb=20 10 600 716,clip]{Cal83c.ps}
}
\caption{\label{f3}}
\end{figure*}

\begin{figure}
\includegraphics[width=\columnwidth, bb=14 50 581 828, clip]{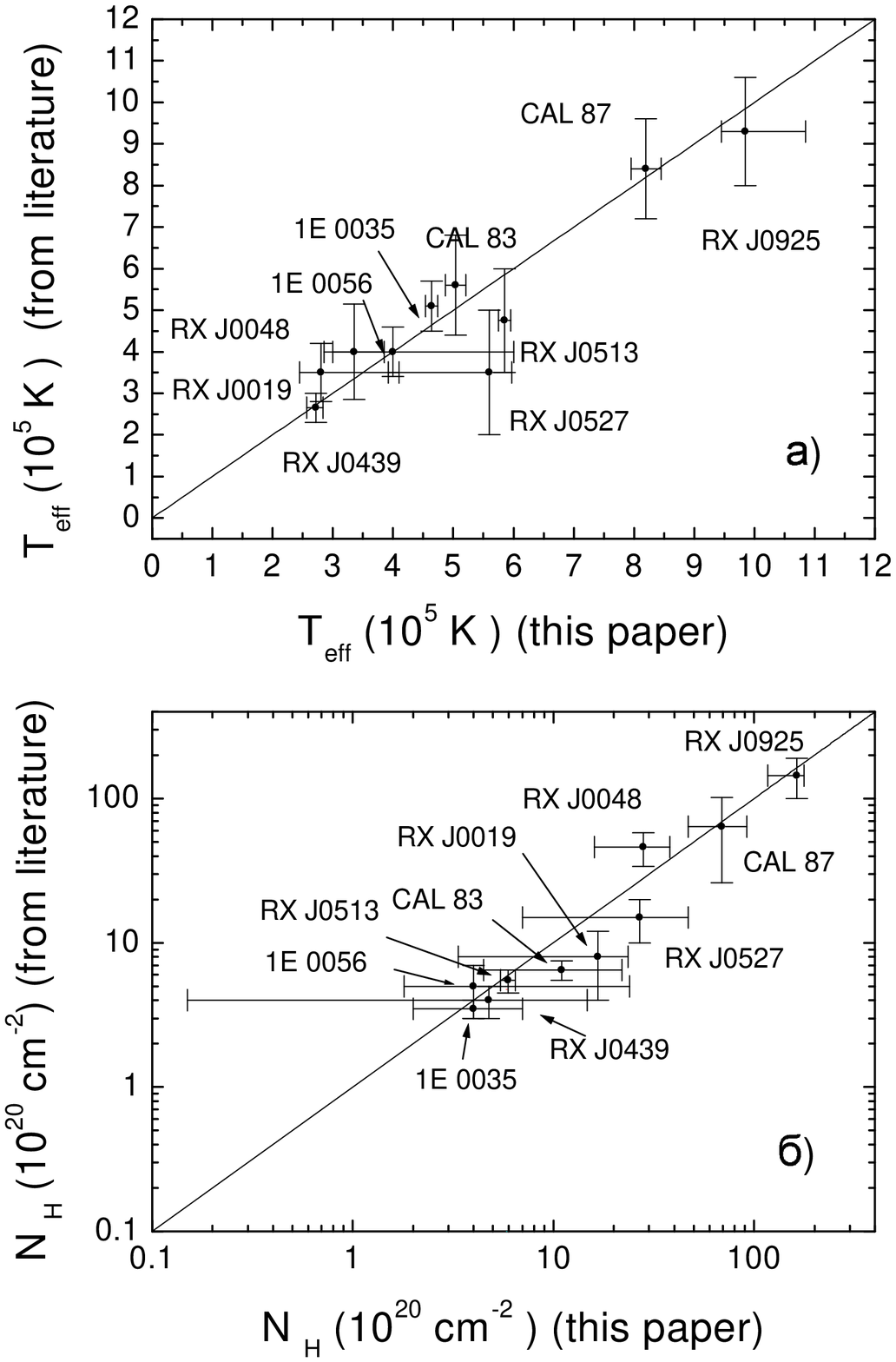}
\caption{\label{fig4}
}
\end{figure}
\begin{figure}
\includegraphics[width=\columnwidth, bb=14 50 581 828, clip]{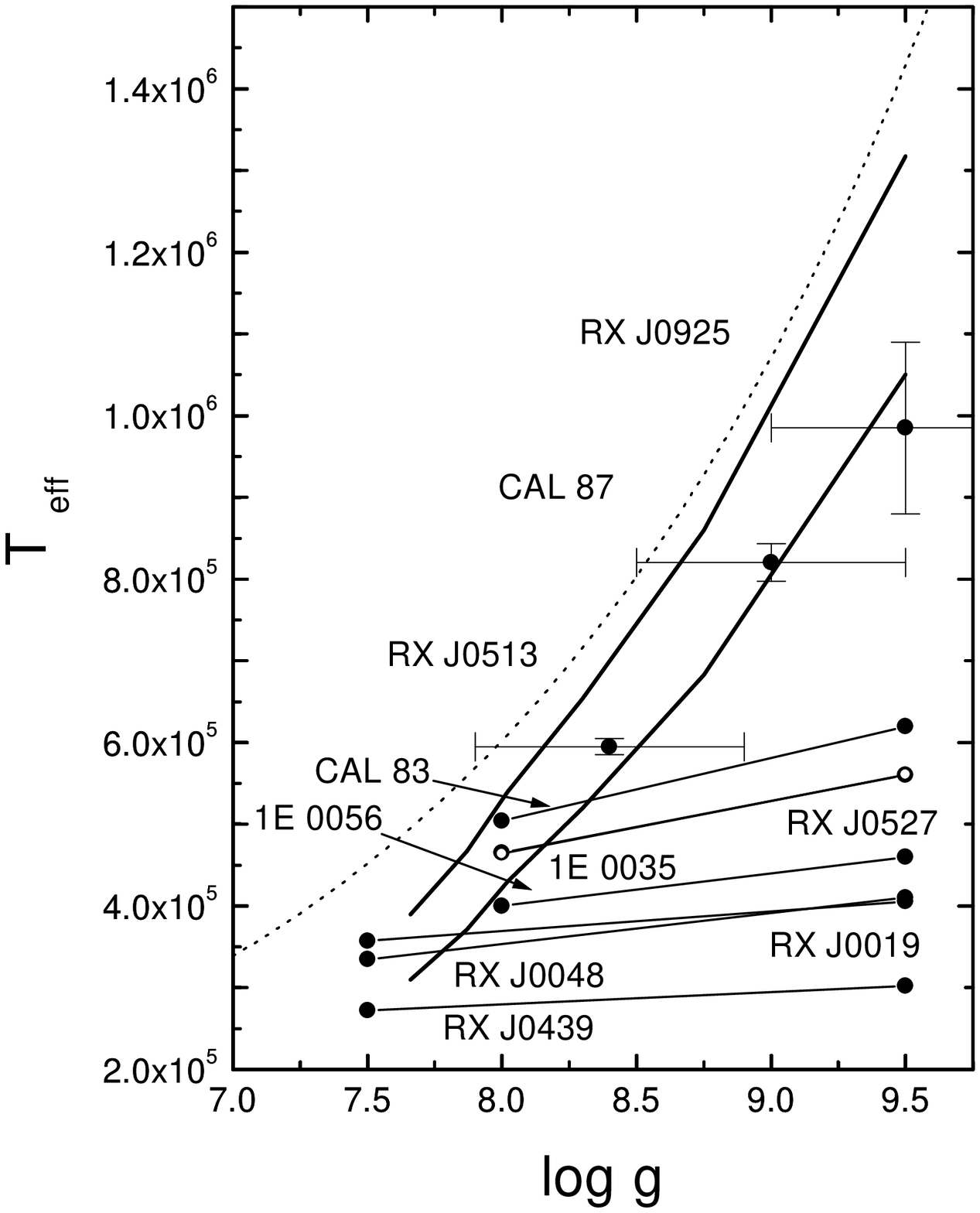}
\caption{\label{fig5}
}
\end{figure}

\end{document}